\documentclass{article}

\usepackage{amssymb}
\usepackage[T1]{fontenc}
\usepackage[left=2.00cm, right=2.00cm, top=1.50cm, bottom=2.00cm]{geometry}
\usepackage{graphicx}
\usepackage{hyperref}
\usepackage[utf8]{inputenc}
\usepackage{latexsym}
\usepackage{natbib}
\usepackage{titling}

\bibpunct{(}{)}{;}{a}{}{,}    
\usepackage[varg]{txfonts}
\graphicspath{{.}}

\usepackage{authblk}

\newsavebox\affbox

\title{\textbf{Ground--based astrometry calibrated by Gaia DR1: new
  perspectives in asteroid orbit determination.}}

\author[1]{F.~Spoto} 
\author[1]{P.~Tanga}
\author[2]{S.~Bouquillon}
\author[3]{J.~Desmars}
\author[3]{D.~Hestroffer}
\author[1]{F.~Mignard}
\author[4,5]{M.~Altmann}
\author[6]{D.~Herald}
\author[7]{J.~Marchant}
\author[2]{C.~Barache}
\author[2]{T.~Carlucci}
\author[8]{T.~Lister}
\author[2]{F.~Taris}

\affil[1]{Université C\^ote d'Azur, Observatoire de la C\^ote d'Azur,
  CNRS, Laboratoire Lagrange, route de l'Observatoire, Nice,
  France}
\affil[2]{SYRTE, Observatoire de Paris, PSL Research University,
  CNRS, Sorbonne Universités, UPMC Univ. Paris 06, 61 avenue de
  l'Observatoire, 75014 Paris, France}
\affil[3]{IMCCE, Observatoire de Paris, PSL Research University,
  CNRS, Sorbonne Universités, UPMC Univ. Paris 06, Univ. Lille, 77
  av. Denfert-Rochereau F-75014 Paris, France}
\affil[4]{SYRTE, Observatoire de Paris, PSL Research University,
  CNRS, Sorbonne Universités, UPMC Univ. Paris 06, 61 avenue de
  l'Observatoire, 75014 Paris, France}
\affil[5]{Zentrum für Astronomie der Universit\"at Heidelberg, ARI,
  Heidelberg, Germany}
\affil[6]{International Occultation Timing Association (IOTA)}
\affil[7]{Liverpool John Moores University, ARI, Liverpool, UK}
\affil[8]{Las Cumbres Observatory, Goleta, USA}

\date{}

\begin{document}
  \maketitle
      {The Gaia Data Release 1 (GDR1) is a first, important step on the path
        of evolution of astrometric accuracy towards a much improved
        situation. Although asteroids are not present in GDR1, this
        intermediate release already impacts asteroid astrometry.}
      {Our goal is to investigate how the GDR1 can change the approach to a
        few typical problems, including the determination of orbits from
        short-arc astrometry, the exploitation of stellar occultations, and
        the impact risk assessment.}
      {We employ optimised asteroid orbit determination tools, and
        study the resulting orbit accuracy and post-fit
        residuals. For this goal, we use selected ground-based asteroid
        astrometry, and occultation events observed in the past. All
        measurements are calibrated by using GDR1 stars.}
      {We show that, by adopting GDR1, very short measurement arcs can
        already provide interesting orbital solutions, capable of correctly
        identifying Near Earth Asteroids (NEAs) and providing a much more
        accurate risk rating. We also demonstrate that occultations,
        previously used to derive asteroid size and shapes, now reach a new
        level of accuracy at which they can be fruitfully used to obtain
        astrometry at the level of accuracy of Gaia star positions.}
      {}
      
      \emph{keywords}: Gaia, asteroids, astrometry, orbit determination, occultations
      

\section{Introduction}  

The ESA Gaia collaboration has published the first Gaia Data
Release (GDR1) containing different data sets of stellar astrometry
\citep{gaia_collaboration_gaia_2016}. Although it is based on a first,
preliminary calibration to be improved by future releases, GDR1
already represents a huge jump in our knowledge of the sky
\citep{gaia_astrometry_2016}. Roughly quantified as a factor $\sim$10
in accuracy, this improvement opens immediate perspectives of
scientific exploitation in many branches of astrophysics.

Gaia is also directly measuring asteroid positions, but they are not
yet present in GDR1 and will appear in the future
intermediate releases (starting with GDR2, April
2018).

While direct measurements of asteroids by Gaia will provide the most
accurate positions available \citep{tanga_gaia_2008}, the improvement
in stellar astrometry itself can potentially change all other
approaches where stars are used as astrometric reference.

In this article, we present the first results obtained in this
respect, by exploiting both recent images, optimally calibrated and
reduced by GDR1, and stellar occultations whose target star is
contained in GDR1.

Our motivation is twofold: 
\begin{itemize}
\item First, Gaia provides an all-sky, dense system of astrometric
  reference sources, that can be used to measure all asteroids, even
  those that are beyond reach for Gaia due to flux limit or geometric
  configuration.
\item Second, this approach extends in time -- beyond the Gaia mission
  duration -- the possibility of measuring asteroid positions at very
  high accuracy by applying a better calibration to old and new data.
\end{itemize}
 
Although the intermediate accuracy of GDR1 is going to be surpassed by
forthcoming releases, the indications that we obtain in this evolving
context are already significant and clearly indicate the change of
paradigm brought by Gaia. For our application, the most relevant
information to come will be parallaxes and proper motions for the
whole sample of $\sim$1 billion stars, that will be made available in
GDR2. Already in GDR1 however, the Tycho-Gaia Astrometric Solution
(TGAS) provides these parameters at very high accuracy for $\sim$2
million stars, exploiting the long time base between Gaia and its
pioneering precursor, Hipparcos/Tycho.

\citet{desmars_statistical_2013} have studied the current situation of
orbital uncertainties for asteroids. Their statistics on the content
of the Minor Planet Center observation database show that most of the
observations are Charge-Coupled Device (CCD) measurements with an
average accuracy of $\sim$400~mas. Optimised surveys can reach much
better accuracy levels, but limitations due to zonal and systematic
errors of the catalogues prevent us from reaching uncertainties better
than $\sim$40~mas \citep{Farnocchia_2015}.

Past studies, such as \citet{Chesley_2010}, \citet{Farnocchia_2015},
have analysed different catalogues to derive local corrections that
can be used to improve the available set of astrometric measurements
archived at the Minor Planet Center. This approach can be applied to
the whole record of the existing astrometry, while the calibration of
the raw data (CCD images, in general) is not re--processed. On the
other hand, our goal hereinafter is to assess the impact of GDR1 at
the very beginning of the process, that is starting from the
calibration procedures.

We consider two data sources that are expected to provide very precise
data.  The first one is the CCD astrometry provided by the
Ground--Based Optical Tracking (GBOT) of Gaia. This activity, which
runs on several telescopes, regularly obtains astrometric data of the
Gaia satellite itself since its launch in Dec. 2013. As the Gaia
trajectory sweeps a relatively large sky area around opposition, in
the direction of L2, GBOT observes and discovers asteroids (typically
between 10 and 80 each night, of which $\sim 46\%$ are new
discoveries). Starting with the availability of GDR1 in September
2016, GBOT has exploited it for the calibration of the
astrometry. Asteroid positions are regularly submitted to the Minor
Planet Center.

The second data source is provided by the data set of stellar
occultations by asteroids, observed in the past. Such events can
provide very accurate positions of the asteroid relative to the
occulted star. The best results are obtained when the sky--projected
shape of the asteroid is sampled by several occultation chords.

We describe in detail our approach in the following sections. GBOT
calibration of asteroid astrometry is illustrated in
Sect.~\ref{S:GBOTcalibration}. We investigate the evolution of
the impact rating for the first confirmed Near Earth Asteroid
NEA discovered by GBOT, using GDR1 in Sect.~\ref{S:NEOrisk}. The
improvement of short--arc orbit determination is then studied in
Sect.~\ref{S:shortarc}. Sect.~\ref{S:occultations} is devoted to the
exploitation of the stellar occultations. Finally our results are
summarised in Sect.~\ref{S:discussion}.

\section{Astrometric calibration of Ground Based Optical Tracking data by Gaia DR1}
\label{S:GBOTcalibration}

GBOT is an observation campaign to organise and carry out the high-
precision astrometric tracking of the Gaia satellite itself, in the
frame of the Data Processing and Analysis Consortium (DPAC) of Gaia.

Its goal is to fully ensure the elimination of systematic effects, for
example, aberration, even for those objects which can be measured to
the greatest precision~\citep{Altmann_2014} and for which traditional
radar tracking methods alone are not fully sufficient.

The GBOT approach is based on daily CCD observations performed
throughout the mission. Currently GBOT mainly uses two telescopes: the
VLT Survey Telescope (VST) installed at ESO's Paranal Observatory in
Chile and the Liverpool Telescope (LT) on La Palma (Canary Islands,
Spain). Each night, each telescope attempts to take a sequence of
frames on which the Gaia satellite itself is seen as a faint and fast
moving object (its magnitude is about R$\sim$21 and its sidereal
speed can reach $40$ mas/s).

The Gaia mission requirement for the absolute accuracy on the
satellite position determination in the International Celestial
Reference Frame (ICRF) is 150 m, which translates to 20 mas in the
plane of sky. This precision cannot be reached by the usual radar
ranging and communications stations, which can only deliver
$\sim$2000~m in position and 10~mm/s in velocity on the sky.  A
specific tool – the GBOT Astrometric Reduction pipeline (GARP) – has
been developed to reach this level of precision and accuracy for the
reduction of images of moving objects – that is, trailed images on the
CCD frame~\citep{Bouquillon_2014}. Note that since the VST tracking is
locked on the Gaia speed, the satellite is recorded as a stationary
object while the stars are recorded as trailed, elongated, images.

Since the beginning of 2015, the GBOT group decided to also measure
with the help of GARP the SSO Solar System Object present in the one
square degree field of VST and in the 10'$\times$10' field of LT. In
April 2017, more than 12000 asteroid positions were recorded and
submitted to the MPC.

GARP proceeds, first, by determining the photo-centre position of the
object as if it was non-moving. The drift angle and amplitude are
determined from the ICRF positions in the first and last frames of the
whole sequence. In a second iteration, a linearly moving Gaussian to
the drifting PSF is fitted~\citep{Bouquillon_2017}.

The precision of this photo-centre determination is the result of
complicated interactions between the brightness of the object, the
elongation due to the drift and the conditions of observation (seeing,
pixel size, background flux, etc.)  (see \cite{Bouquillon_2017} for
more details). For a bright and slowly moving Main belt asteroid such
as (1132) Hollandia (R $\sim14.5$ mag, S/N $\sim 800$, elongation
$\sim2$ Pixels), which is one of the brighter MBAs found in the GBOT
data, the uncertainty is around 5 mas for an exposure time of one
minute. For a faint and fast object observed in similar conditions,
such as the NEA $2016~EK_{85}$ (R $\sim 20$ mag, S/N $\sim
10$,elongation $\sim 21$ Pixels), the centroiding error can reach 100
mas of which the largest part is along the drift.

For the astrometric calibration, we use a reference catalogue as a
realisation of the ICRF. For each image, we compute the standard
coordinates of all reference stars potentially in the field of view by
taking into account the aberration, the tropospheric effect and by
applying a gnomonic projection. Then an algorithm based on the planar
triangle method~\citep{Liebe_1995} makes the connection between each
reference star and its counterpart in the image. Finally, two
polynomials are fitted to perform the link between the X and Y
coordinates of the CCD frame and the standard coordinates. For
OMEGACAM at VST a polynomial of first degree is sufficient,
while three degrees are necessary for the IO:O camera at LT.

\begin{figure*}[t]
 \begin{center}
  \includegraphics[width=16 cm]{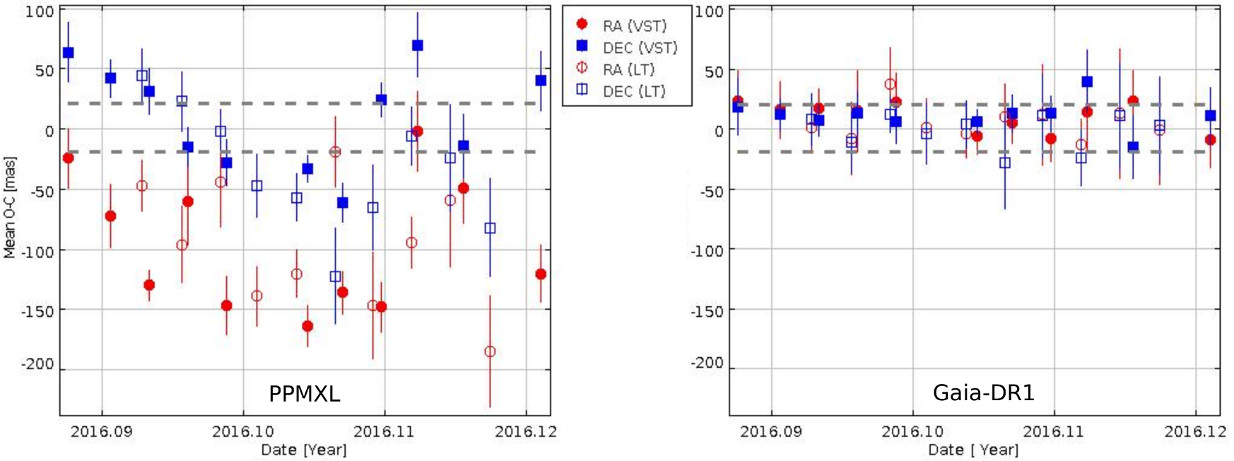}
  \caption{ Residuals between predicted and measured positions of the
    Gaia satellite. The astrometry has been calibrated with PPMXL
    (left panel) and Gaia DR1 (right panel). Each symbol corresponds
    to the average over a sequence of several frames over 30 minutes
    maximum (VST: filled symbols; LT: open symbols). Squares and
    circles indicate residuals in right ascension and declination
    respectively.}\label{fig:GARP_Gaia}
 \end{center}
\end{figure*}

Using Gaia DR1 (Gaia consortium, 2016) as the reference catalogue, the
standard deviation of the residuals between the catalogue and the data
for all stars with V$<$20.5 mag is $\sim$30 mas. This level of
precision is not attainable without using Gaia astrometry as a
reference catalogue. For instance, if we follow the same procedure of
calibration with the PPMXL stars catalogue~\citep{Roeser_2010}, the
standard deviation is $\sim$300 mas, that is ten times larger
(see Fig.~\ref{fig:GARP_Gaia}).

A second advantage of using Gaia DR1 as a reference catalogue, is to
eliminate zonal errors. To illustrate this point,
Fig.~\ref{fig:GARP_Gaia} presents the results of the GARP reductions
of 21 sets of observations of the Gaia satellite itself, covering 13
consecutive nights in February 2016. Note that since the angular speed
of Gaia is one degree per day, the fields of view of two sets of
observations taken at a 24 hours interval are completely different. In
the case of the reduction with PPMXL (left panel), we observe a
progressive variation of the daily mean differences between the Gaia
position measured on the CCD frames and its ephemeris. Over a period
of 13 days, these variations reach 150 mas in right ascension and
around 100 mas in declination. These amplitudes are several times
larger than the precision (around 30 mas) of each mean measurement in
the reference frame of the PPMXL catalogue. This is confirmed by the
calibration with Gaia DR1 (right panel), showing residuals below 20
mas, in good agreement with the ephemeris.

\section{NEOs and risk rating}
\label{S:NEOrisk}

The first confirmed NEA discovered by GBOT is the asteroid
$2016~EK_{85}$. This object represents a nice example illustrating how
the accuracy of the astrometric calibration can affect the
interpretation of short--arc orbital solutions in the frame of
NEA impact monitoring.

$2016~EK_{85}$ was observed for the first time the night of March 9,
2016, by VST. It was then re--observed
the following night by LT, and other
observers.

The Minor Planet Center classified the newly discovered object as a
NEA and published the corresponding Minor Planet Electronic Circular
(MPEC)\footnote{\url{http://www.minorplanetcenter.net/mpec/K16/K16EC2.html}},
with $48$ observations ($28$ GBOT and $20$ from other
observers). After the discovery of a new NEA (and whenever new
observations are added), both NEODyS in
Pisa\footnote{\url{http://newton.dm.unipi.it/neodys2/}} and Sentry at
the JPL\footnote{\url{http://neo.jpl.nasa.gov/risks/}} check the
possibility of impacts with the Earth for $100$ years in the future.

At that early stage after the discovery, it turned out that the object
had predicted possible impacts with the Earth in $2102$ and $2106$,
with low impact probability (IP,
Table~\ref{tab:2016EK85_risk}).

One week later, on March 16, new observations at Mauna Kea (published
on MPEC
2016-F48\footnote{\url{http://www.minorplanetcenter.net/mpec/K16/K16F48.html}})
produced a better orbital solution, definitely ruling out the
possibility of an impact.

\begin{table}[h]
  \begin{center}
    \caption{Date, impact probability and Palermo Scale rating of possible
      impacts with the Earth for the NEA $2016~EK_{85}$. The impact
      table appeared in NEODyS on the night of March 11,
      2016. It has been computed using $48$ optical observations from
      March 9 to March 11 by the OrbFit software version 5.0. All
      these data are consistent with data published by the
      JPL.}\label{tab:2016EK85_risk}
    \begin{tabular}{ccc}
      \hline
      Date & IP & PS \\
      \hline
      2102/02/22.296   &  $1.24 \ 10^{-8}$  & $-8.54$\\
      2102/02/22.549   &  $5.57 \ 10^{-8}$  & $-7.89$\\
      2106/02/22.042   &  $6.82 \ 10^{-8}$  & $-7.82$\\
      2106/02/22.311   &  $4.62 \ 10^{-7}$  & $-6.99$\\
      2106/02/22.529   &  $1.20 \ 10^{-6}$  & $-6.57$\\
      2106/02/22.605   &  $3.19 \ 10^{-7}$  & $-7.14$\\
      2106/02/22.635   &  $8.93 \ 10^{-6}$  & $-5.70$\\
      \hline
    \end{tabular}
  \end{center}
\end{table}

When $2016~EK_{85}$ was observed in March 2016, GDR1 was not yet
available. We thus decided to reprocess the whole set of 28 GBOT
observations covering two nights (2016/03/09 and 2016/03/10), by
exploiting the calibration based on GDR1. The sample contains $8$
positions from the VST and $20$ from the LT. Then we add to these $28$
observations, the other $20$ observations to reproduce the same
initial set.

A preliminary orbit was computed with the Gauss method. The final
orbit is obtained after a weighted least squares fit with an outlier--
rejection procedure \citep{Carpino_2003}. In principle, orbital
fitting would require a weighting scheme based on an independent
assessment of the accuracy obtained on asteroids by the application of
different catalogues \citep{Farnocchia_2015}. In the case of GDR1 we
searched for the appropriate weights consistent with the post-fit
residuals.

\begin{figure}[h]
  \centering
  \includegraphics[width=8 cm]{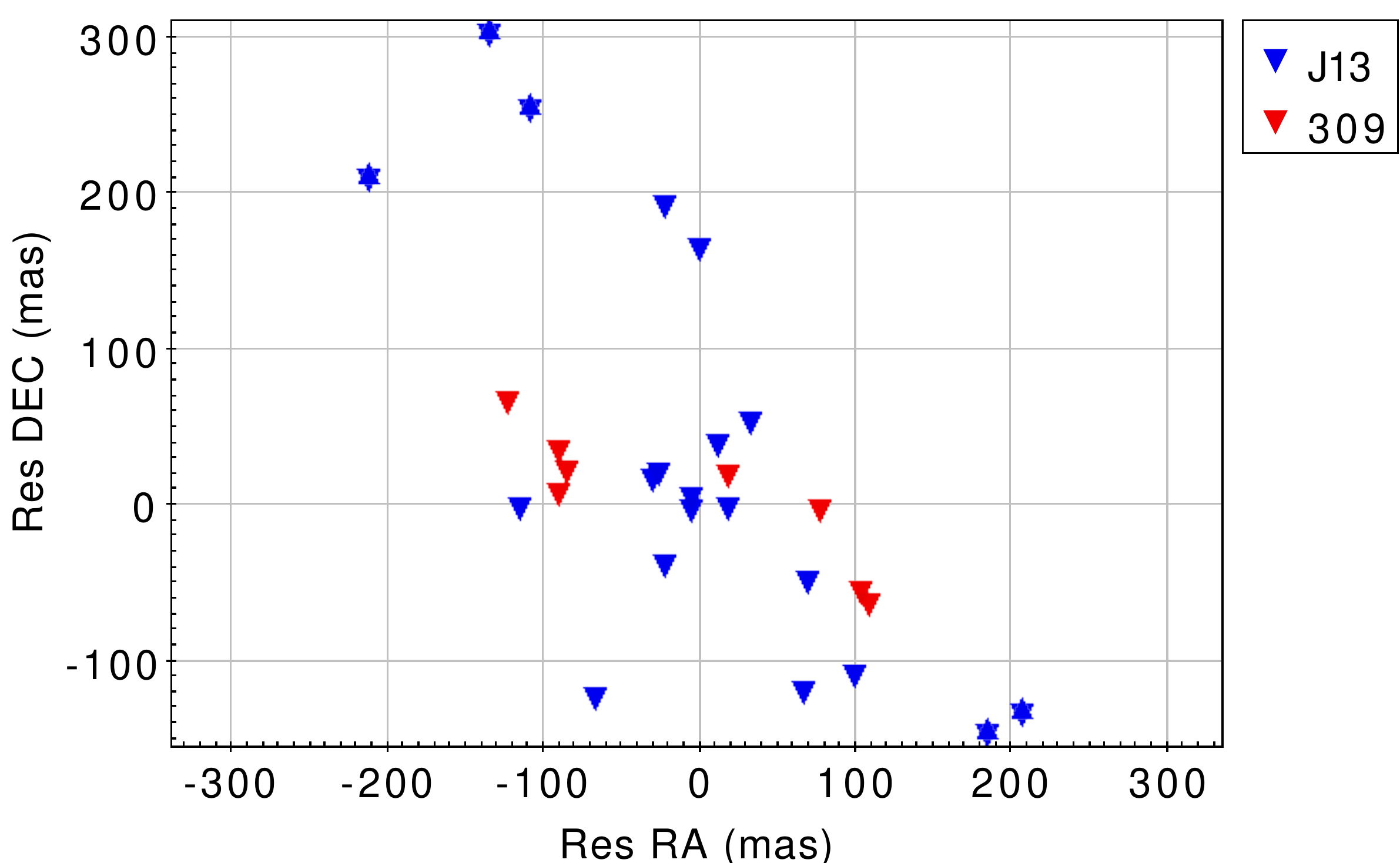}
  \caption{Residuals in right ascension (RA) and declination (DEC) for
    the GBOT Gaia reduced observations. Stars represent the outliers
    discarded by the outlier--rejection procedure. The residuals are
    considered separately for the VST (MPC code 309) and the LT (MPC
    code J13).}\label{fig:2016EK85_res}
\end{figure}

Figure~\ref{fig:2016EK85_res} shows the final residuals, that are
typical of what we expect in GBOT for very faint moving sources. Few
observations are discarded by the rejection procedure.

We consider three different orbits, one (\emph{Orbit 1}) obtained
using the initial set of observations as submitted at the MPC, another
(\emph{Orbit 2}) with the same set of observations, but where GBOT
observations are reduced using GDR1, and the final orbit (\emph{Final
  Orbit}) obtained using all the observations available at the MPC. We
have compared \emph{Orbit 1} and \emph{Orbit 2} to \emph{Final orbit}
using different metrics. The result is that \emph{Orbit 2} is closer
to \emph{Final orbit} than \emph{Orbit 1}. This is of course due to
the quality of the Gaia catalogue and to the consequent better
reduction of GBOT observations.

We have used two different metrics to compare the orbits. The first
one ($d$) is the most simple, and it represents the difference of the
Equinoctial orbital elements:
\begin{equation}
   d = \sqrt {\left ( \frac{(a_1-a_2)}{(a_1+a_2)} \right )^2 + (h_1-h_2)^2 + (k_1-k_2)^2 + (p_1-p_2)^2 + (q_1-q_2)^2} 
\end{equation}
\normalsize 
The second one ($d_{LoV}$) is based on the orbit identification
algorithm, as described in~\cite{Milani_2005_metrics}. We use two sets
of virtual asteroids~\citep{Milani_2005}, and we compute the expected
$\chi^2$ for the identification between each pair of virtual
asteroids. Then we select the minimum value to obtain the final
comparison. This identification algorithm is very useful when
non-linearity plays an essential role, as in this case in which the
time span by the observations in
short. Table~\ref{tab:2016EK85_metrics} summarises the results of the
comparison, and shows that using both metrics \emph{Orbit 2} (obtained
using GDR1) is always closer to the final orbit.

\begin{table}[h]
  \begin{center}
    \caption{\emph{Orbit 1} and \emph{Orbit 2} compared to the final
      orbit available through two different metrics: $d$ is the
      difference in Equinoctial elements, and $d_{LoV}$ is based on
      the identification algorithm applied to the Line of
      Variations~\citep{Milani_2005_metrics}.}\label{tab:2016EK85_metrics}
    \begin{tabular}{ccc}
      \hline
      Orbits to be compared & $d$ & $d_{LoV}$ \\
      \hline
      \emph{Orbit 1 - Final orbit} & $0.0010$ & 1.20 \\
      \emph{Orbit 2 - Final orbit} & $0.0004$ & 0.01 \\
      \hline
    \end{tabular}
  \end{center}
\end{table}

Based on the obtained orbit, we have looked for possible impacts with
the Earth in the next 100 years, as was done when the object was
studied for the first time. We computed multiple solutions, and
analysed each close encounter and each return that could lead to a
possible impact. In this case, we do not find any possible impact
risk. This result corresponds to our expectations, as a change in the
astrometry due to an improved catalogue changes the direction of the
Line of Variations (LoV,~\cite{Milani_2005}). This rules out
all the possible impacts that were found with the preliminary solution
based on a less precise catalogue.

\begin{figure}[h]
  \centering
  \hspace{-0.9cm}
  \begin{minipage}{0.2\textwidth}
    \includegraphics[width=3 cm]{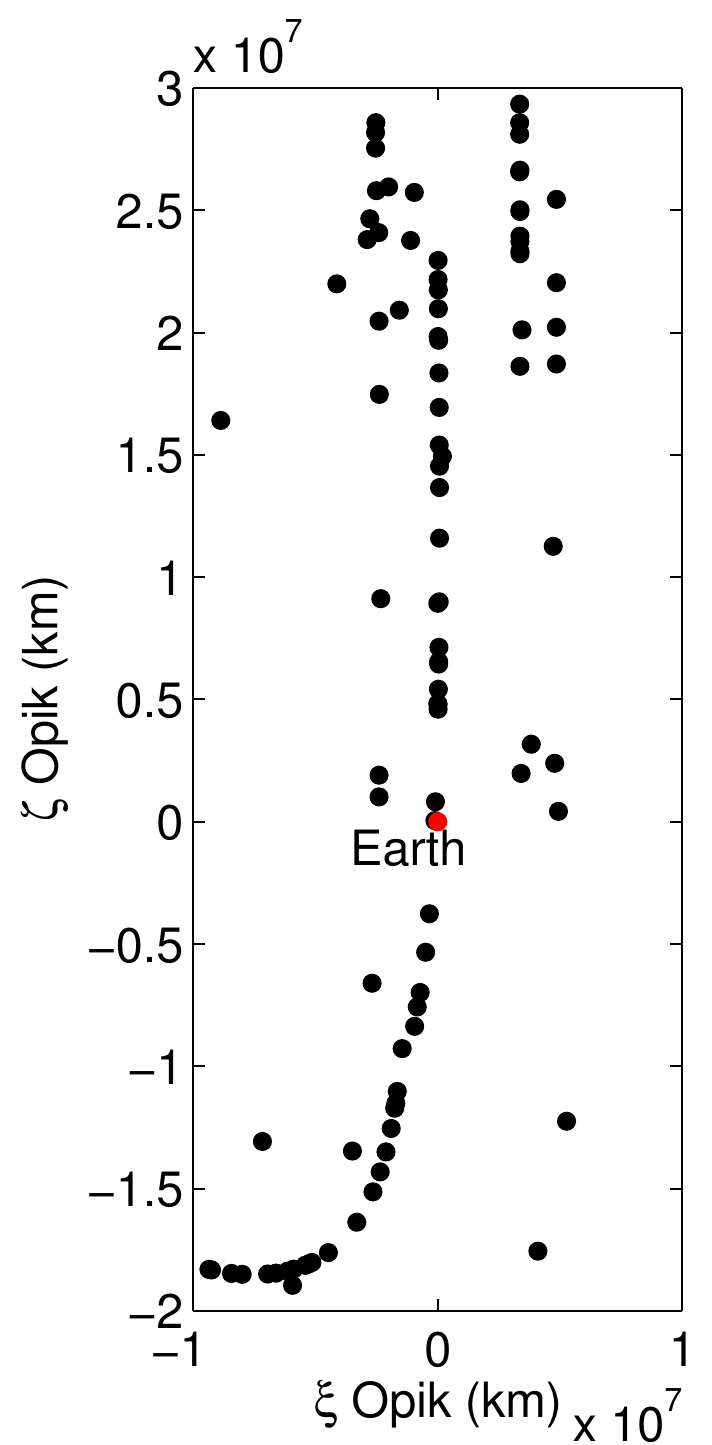}
  \end{minipage}
  \hspace{0.4cm}
  \begin{minipage}{0.2\textwidth}
    \includegraphics[width=4.95 cm]{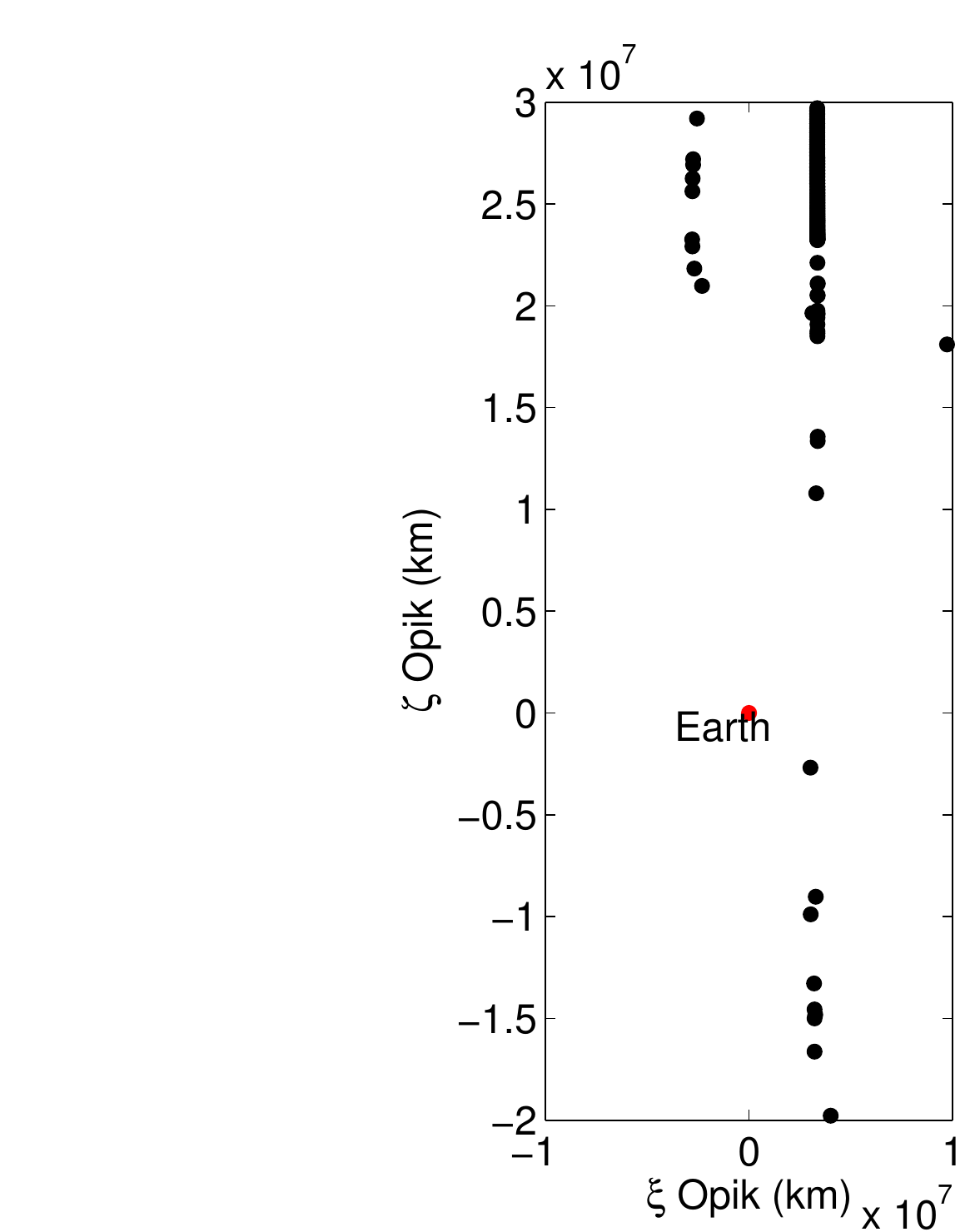}
  \end{minipage}
  \caption{Left: Line of Variations in the Target Plane of 2102 with
    GBOT observations as in the MPC observations file . Right: Line of
    Variations in the Target Plane of 2102 with GBOT observations
    reduced with the GDR1.}\label{fig:LOV}
\end{figure}

Figure~\ref{fig:LOV} shows the LoV using the same set of observations
(GBOT): the left panel reproduces the situation as it was at the
beginning when $2016~EK_{85}$ was put on the risk list, while the
right panel shows what happens using observations reduced by GDR1. It
is clear that in the first case the LoV pass through the Earth, while
in the second it is clearly displaced. The LoV behaviour is very
similar during the close encounter of $2106$ as well.

\section{Exploitation of short observational arcs}
\label{S:shortarc}

Besides the case of $2016~EK_{85}$ we decided to analyse other NEAs or
MBAs observed by GBOT, to explore a wide variety of situations: from
objects with few observations covering a very short time span of
the order of several minutes, to others observed over one or
two nights.

When we have too few observations or too short time span we encounter
the worst possible scenario for the orbit determination: it is not
possible to compute a preliminary orbit, neither to apply a least
squares fit.

We then apply the systematic ranging \citep{Farnocchia2015_Systematic,
  Spoto2017}. This technique allows us to scan the admissible region
\citep{Milani_2005_admissible} using a grid in the plane defined by
topocentric distance and topocentric velocity. The systematic ranging
is used when the amount of information in the observed arc is too
limited to compute a six parameter orbit, and the differential
correction procedure fails. Our goal is thus simply to provide a first
constraint on the family of orbits that are compatible with the
astrometry, and try to distinguish NEAs from other categories.

\begin{figure}[h]
  \centering
  \includegraphics[width=8 cm]{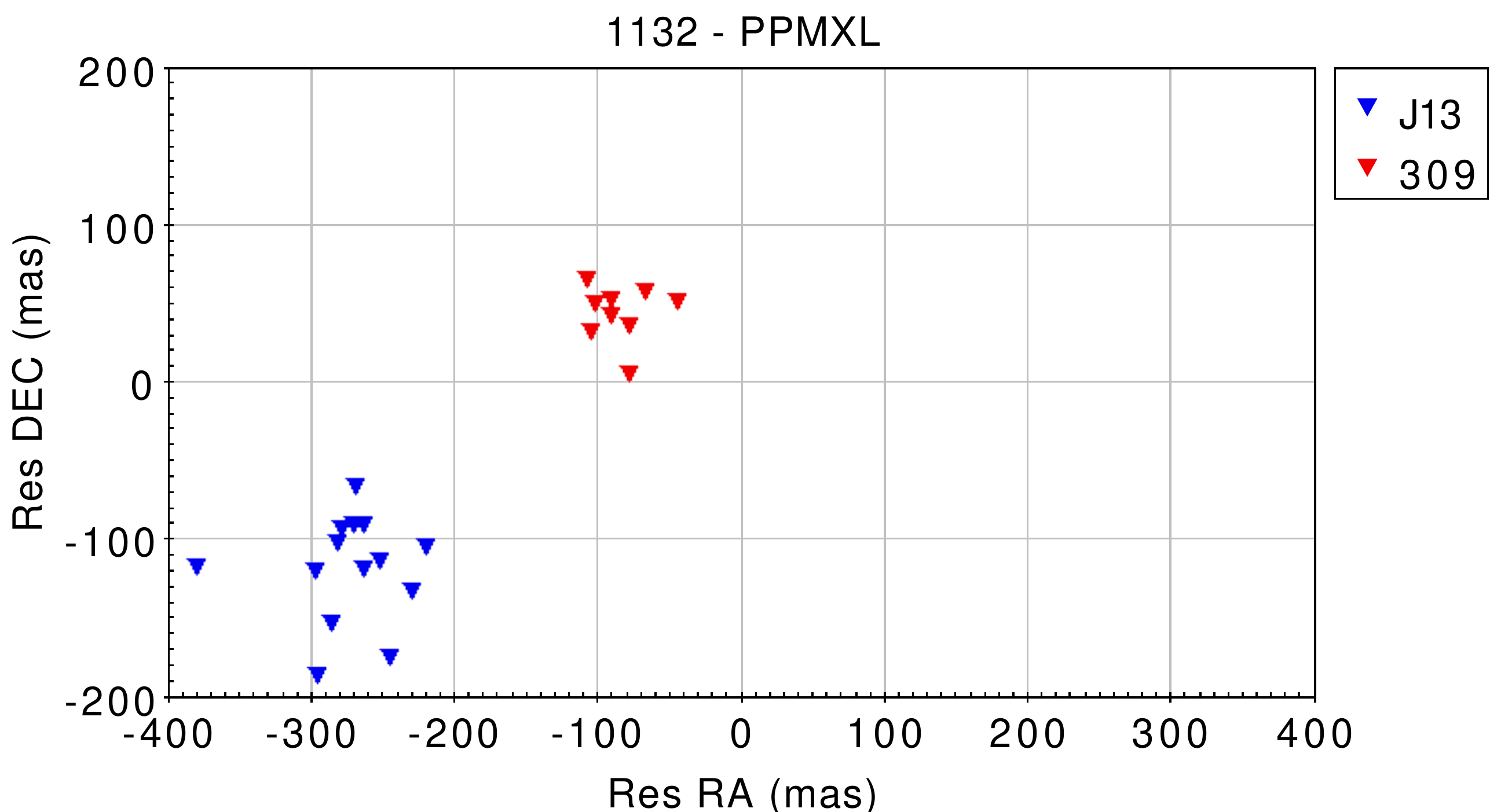}
  \includegraphics[width=8 cm]{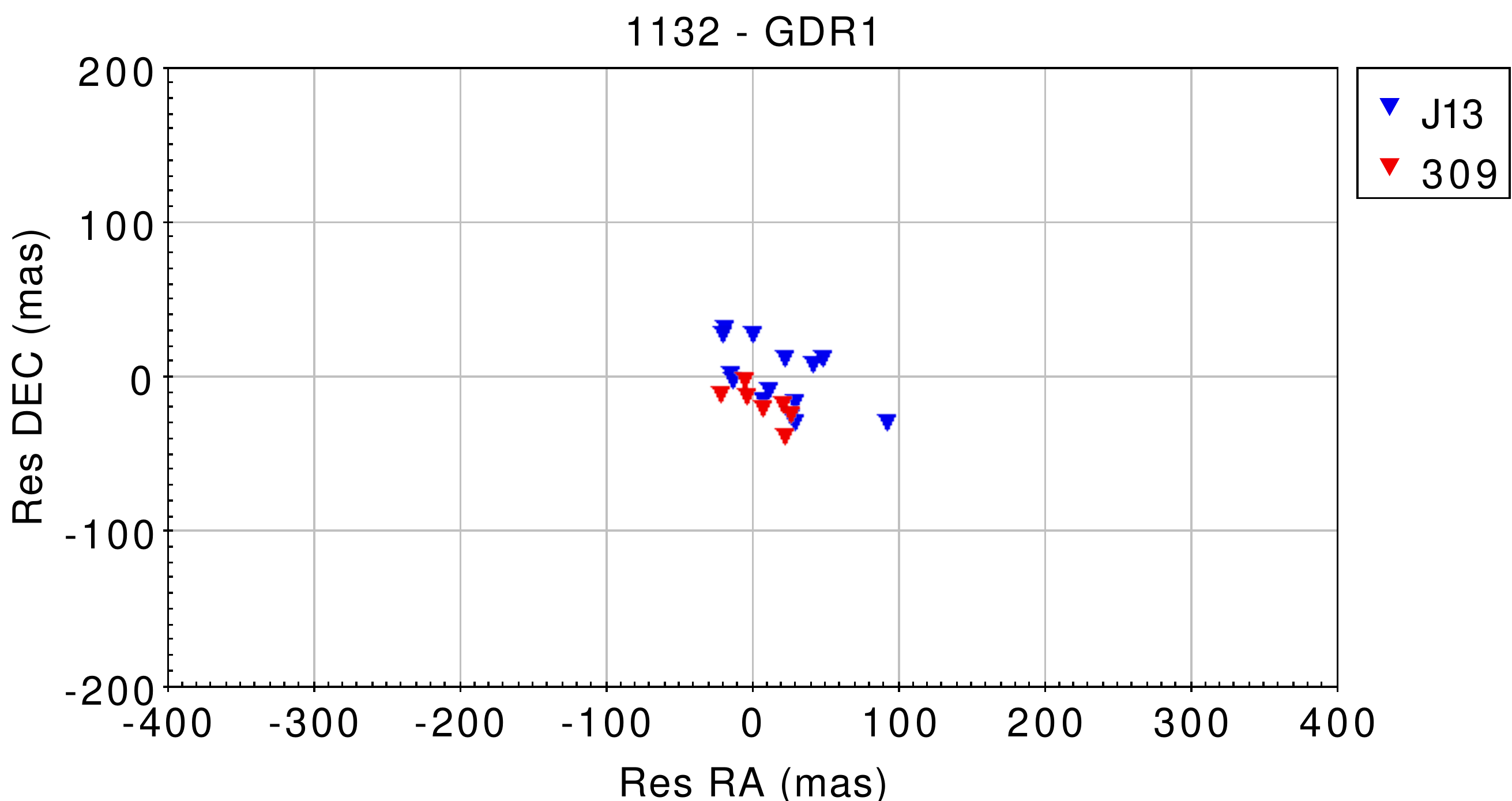}
  \caption{Residuals in right ascension and declination for the main
    belt $(1132)$ Hollandia of GBOT observations, reduced with PPMXL
    (upper panel) and GDR1 (bottom panel). In both cases, the
    residuals are considered separately for the VST (MPC code $309$)
    and the LT (MPC code $J13$).}\label{fig:1132_res}
\end{figure}

As already pointed out in Sect.~\ref{S:NEOrisk}, the choice of the
error model is crucial when we try to determine and fit an orbit, even
when we apply the systematic ranging. Since we still do not have an
error model that could represent the reduction with GDR1, we combine
GBOT observations reduced with GDR1 with the whole set of ground-based
observations available. We then fit the orbit and analyse GBOT
residuals. For one case, namely $(1132)$ Hollandia (already cited in
Sect.~\ref{S:GBOTcalibration}), we have also compared the residuals
obtained reducing GBOT observations with PPMXL and GDR1,
respectively. Figure~\ref{fig:1132_res} shows the residuals of GBOT
observations in right ascension and declination for the main belt
$(1132)$ Hollandia, reduced with PPMXL and GDR1 respectively. The
reduction with the PPMXL catalogue clearly presents some zonal errors
that are completely removed using the Gaia catalogue.

We have computed the mean and the standard deviation of the residuals
for each object analysed (see Table~\ref{tab:res_GBOT}), and then we
have applied the systematic ranging.

\begin{table*}[b]
  \begin{center}
    \caption{Asteroid number, provisional designation, weights in right ascension
      (RA) and declination (DEC) for the two main GBOT observatories:
      LT (J13) and VST (309)}\label{tab:res_GBOT}
    \begin{tabular}{rlrrrr}
      \hline
      Number   & Provisional    &  J13 RA   & J13 DEC  & 309 RA  & 309 DEC \\
               & designation    &  (mas)    & (mas)    & (mas)   &   (mas)  \\
      \hline
      $18109$  & $2000~NG_{11}$   &    &    & 60 & 50\\
      $21565$  & $1998~QZ_{102}$  & 50 & 70 & 30 & 20\\
      $32998$  & $1997~CK_5$     & 30 & 20 & 20 & 30\\
      $79134$  & $1990~VO_8$     & 30 & 35 & 32 & 37\\
      $110520$ & $2001~TL_{79}$   & 23 & 34 & 24 & 30\\
      $119177$ & $2001~QN_{61}$   & 37 & 31 & 34 & 26\\
      $135121$ & $2001~QO_{137}$  & 43 & 30 & 30 & 30\\
      $139832$ & $2001~RL_{35}$   & 43 & 20 & 25 & 23\\
      $162000$ & $1990~OS$       &    &    & 40 & 50\\
      $186822$ & $2004~FE_{31}$   &    &    & 40 & 40\\
      $190788$ & $2001~RT_{17}$   &    &    & 66 & 70\\
      $307301$ & $2002QG_{20}$    & 60 & 50 & 34 & 33\\
      $392704$ & $2012~AE_1$     &    &    & 100 & 100\\
      \hline
    \end{tabular}
  \end{center}
\end{table*}

For the situations that we analyse, only the astrometric
reduction by GDR1 produces exploitable results. We tested our approach
on a set of NEAs and MBAs. We expect, for NEAs, to consistently obtain
a high probability of being a genuine NEA, and that there is no
confusion between the two categories (i.e. one should not find a MBA
with a high NEA score).

Tables~\ref{tab:NEA_syst_rang} and \ref{tab:MBA_syst_rang}
summarise our results. All the objects have a small number of
observations, and the time span is very short, usually less than $20$
minutes for NEAs and $4$ hours for MBAs. The results are perfectly
consistent, and confirm that GDR1 allows us to classify asteroid
orbits, and identify new NEAs by exploiting ground-based astrometry
on a very short observational arc.

\begin{table*}[b]
  \begin{center}
    \caption{Near-Earth asteroid number, provisional designation,
      total number of observations, time span covered (minutes), and
      probability of being a NEA.}\label{tab:NEA_syst_rang}
    \begin{tabular}{rlrcr}
      \hline
      Number   & Provisional    & No. obs & Time span & Probability   \\
               & designation    &         &   (min)   &  to be NEA     \\
      \hline
      $18109$  & $2000~NG_{11}$  &    8    &   16      & 82\\
      $162000$ & $1990~OS$      &   10    &   16      & 89\\
      $186822$ & $2004~FE_{31}$  &   10    &   16      & 86\\
      $190788$ & $2001~RT_{17}$  &   10    &   16      & 100\\
      $392704$ & $2012~AE_1$    &    6    &   14      & 90 \\
      \hline
    \end{tabular}
  \end{center}
\end{table*}

\begin{table*}[b]
  \begin{center}
    \caption{Main Belt asteroid number, provisional designation, total
      number of observations, time span covered (hours), and probability
      of being a MBA}\label{tab:MBA_syst_rang}
    \begin{tabular}{rlrcr}
      \hline
      Number   & Provisional    & No. obs & Time span & Probability  \\
               & designation    &         &  (hours) &  to be MBA   \\
      \hline
      $21565$  & $1998~QZ_{102}$  &   30    &   3.7    &  96        \\
      $32998$  & $1997~CK_5$     &   30    &   3.0    &  100       \\
      $79134$  & $1990~VO_8$     &   29    &   3.5    &  90        \\
      $110520$ & $2001~TL_{79}$   &   15    &   3.0    &  83        \\
      $119177$ & $2001~QN_{61}$   &   28    &   3.6    &  80        \\
      $135121$ & $2001~QO_{137}$  &   30    &   3.5    &  99        \\
      $139832$ & $2001~RL_{35}$   &   30    &   3.4    &  95        \\
      $307301$ & $2002QG_{20}$    &   30    &   3.3    &  100       \\
      \hline
    \end{tabular}
  \end{center}
\end{table*}

\section{Astrometry by stellar occultations}
\label{S:occultations}

Improved orbits obtained by direct asteroid astrometry by Gaia, and
improved stellar positions, are expected to strongly expand the number
of stellar occultation predictions that have a good probability of
success \citep{tanga_asteroid_2007}. This will provide obvious benefits
for our capability to determine precise asteroid sizes. Also, asteroid
shapes and satellite systems will be better constrained by a large
amount of successful occultations.

Besides this promising perspective, we cannot neglect the value that
the current record of positive stellar occultations has. In fact,
stars contained in GDR1 that have been occulted by asteroids in the
past correspond to very precise astrometric positions at the
corresponding occultation epoch.

A complete list of occultation results, maintained by Dave Herald, is
at the base of an available data set of observed events
\citep{pds_occultations_2016}. Most event epochs are after the year
2000, but some events have been observed starting from the late
70s. The earliest isolated occultation observed dates back to 1961,
for the asteroid (2) Pallas. The database above contains
identifications of the occulted star, of the occulting asteroid, and
the occultation parameters derived from the observations. Such
parameters are computed for the geocentre, and include the apparent
distance between the star and the asteroid at the minimum separation,
and the epoch at which the minimum separation occurs. Both quantities
are listed with their uncertainties, that we discuss in the following
section.

\subsection{Accuracy budget}
\label{S:accuracy}

The occultation accuracy can approach that of GDR1 for the stars,
that is one or two orders of magnitude better than traditional
small-field imaging astrometry of asteroids from the ground. This is
especially true if we consider that most of the occulted stars
observed up to now have magnitudes $V\sim$9-12 mag, a range in which
the Gaia position accuracy is very high.

More precisely, the uncertainty of asteroid astrometry as derived by
occultations is due to:
\begin{itemize}

\item Timing uncertainty and timing errors. Starting from the 90s
  electronic timing synced to GPS has been increasingly adopted, but older
  events were sometimes timed by eye and manual chronometer, thus
  inducing a potentially larger uncertainty subject to personal
  equations. Occasional errors have been detected also on GPS timings. When
  no bias is introduced by technical issues, the uncertainty is
  essentially dominated by the duration of the single exposures used
  to sample the stellar flux. The timing error is translated to an
  uncertainty on the position of the asteroid along the direction of
  its apparent motion.

\item The uncertainty of the relative star-asteroid distance
  transverse to apparent motion, at point of closest appulse. 
  This uncertainty can vary in a wide range, from the
  order of the apparent asteroid size (when only one occultation chord
  is observed) to a very small fraction of it (when several
  occultation chords are measured and the asteroid profile is well
  resolved).

\item The limited accuracy on the position of the star. This is
  strictly dependent on the properties of the stellar catalogue used
  to reduce the observation, and it is the main factor impacted by the
  GDR1/TGAS release.

\end{itemize}

\begin{figure}[h]
  \centering
  \includegraphics[width=7 cm]{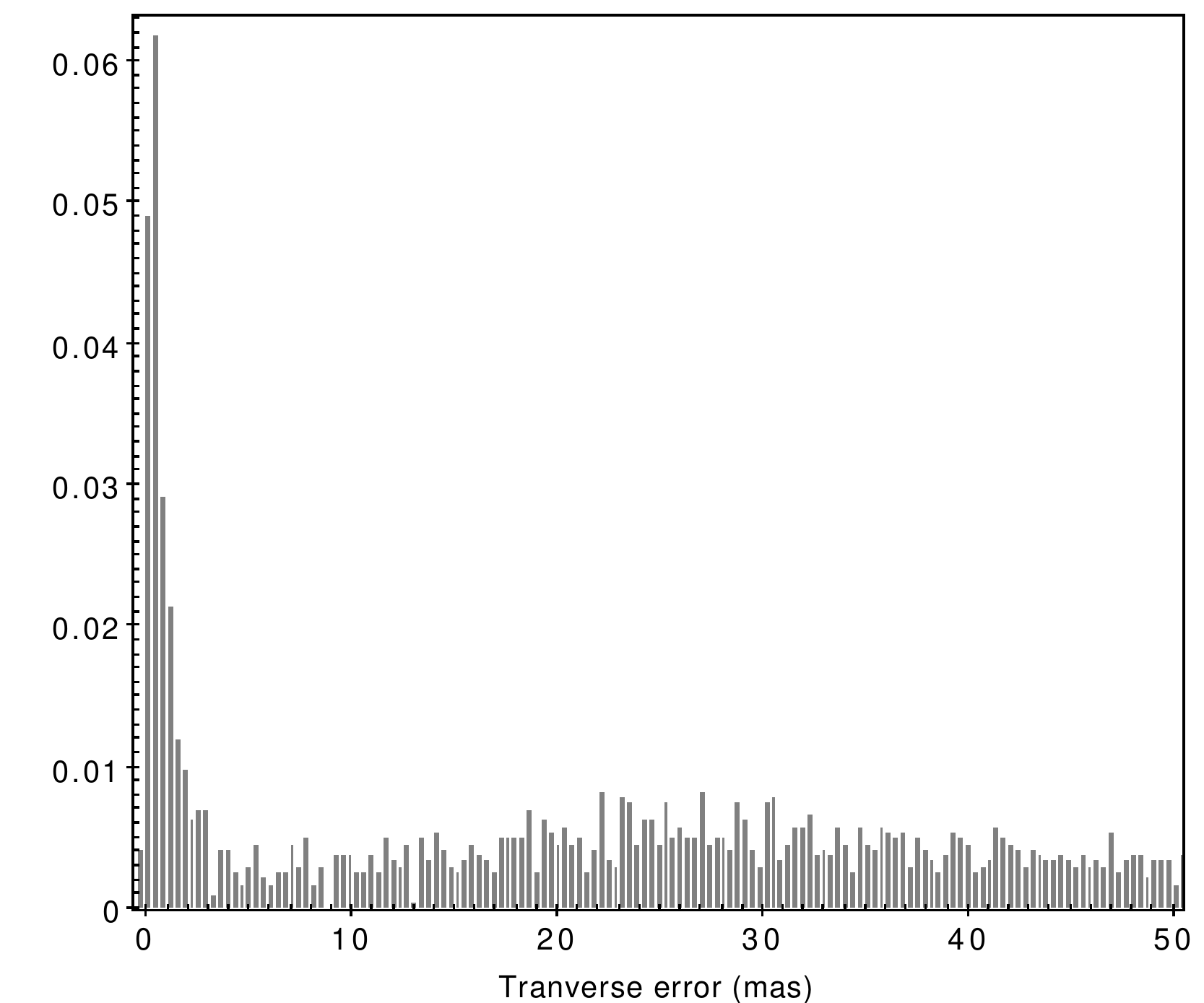}
  \caption{Histogram of the astrometric error on the asteroid position
    in the direction perpendicular to the apparent motion of the
    asteroid during the occultation event.}\label{fig:transverse_err}
\end{figure}

The occultation database provides errors in both timings and
transverse direction (the first two components above). Concerning
timings, a large range of uncertainty values is present, but a clear
peak around $\sim$0.05~s appears. For a typical main belt apparent
motion of 10-15~mas/s, this corresponds to an error around 0.5-1~mas.

Concerning transverse errors, the distribution is clearly bi-modal. A
first peak for small values ($\sim$ 1~mas) corresponds to the best
observed, multi--chord events. A second, much more spread-out set of
values, has a wide, flat maximum in the range 20-40~mas (see
Fig.~\ref{fig:transverse_err}). This is the same order of the typical
apparent radius of the occulting asteroids.

For this first attempt of exploitation of occultation astrometry, we
consider only the transverse uncertainty, as, in general, it is larger
and dominates the error budget. This is not really a limitation at
present, as an accurate use of the timing information is much more
delicate and could require a detailed check of those occultation
events that could be affected by timing anomalies.

By being conservative and using a single uncertainty value, we believe
that we incorporate most error sources in our bulk exploitation of
occultation data, without optimistic assumptions. In a forthcoming
work we will consider a more detailed analysis of selected asteroids
and events.

Concerning the occulted stars we apply the following approach:
\begin{itemize}
\item We match the position of the star to the GDR1, looking for the
  position of corresponding sources in a 2 arcsec radius; we then
  apply a further check on the consistency of the magnitude (which is
  in fact redundant, as no ambiguities are found). The position of the
  matched source in GDR1 is adopted for the occultation. For stars not
  present in GDR1, the corresponding events are discarded.

\item If the star is in TGAS, the position is corrected for
  its proper motion, consistently with the time delay between
  the observed event and the catalogue epoch.

\item If a star is not in TGAS, it does not have a proper motion. In
  this case we consider the difference in position between the UCAC4
  position and GDR1 to compute an approximate proper motion.
\end{itemize}

This approach has clear limitations for non-TGAS sources (mainly due
to the zonal error in UCAC4) and further tests of GDR1 against other
astrometric catalogues could provide useful information; we
consider that this investigation, however, is beyond our immediate goals of
globally testing the approach on the whole set of occultations.

\subsection{Orbit adjustement}
\label{S:occorbits}
 
We attempted an orbital determination, using occultation astrometry
alone, for all asteroids that have a historical record of more than four
occultations. The observation weights are represented by the
transverse accuracy, as explained above. Our fitting procedure rejects
observations whose residual is incompatible with the
weight~\citep{Carpino_2003}. No other observations are included for
the moment in the orbit fitting procedure.

\begin{figure}[h!]
  \centering
  \includegraphics[width=10 cm]{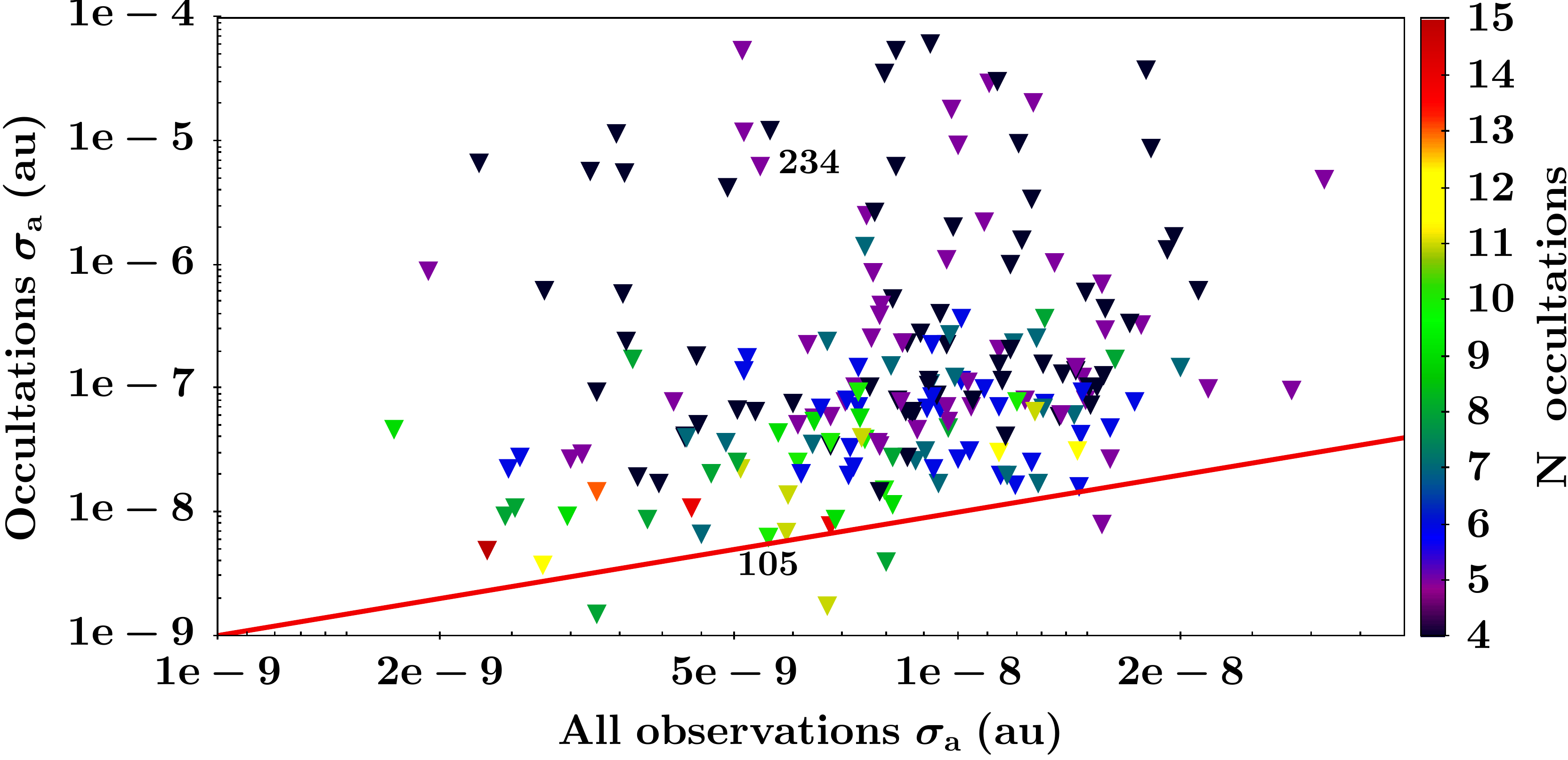}
  \includegraphics[width=10 cm]{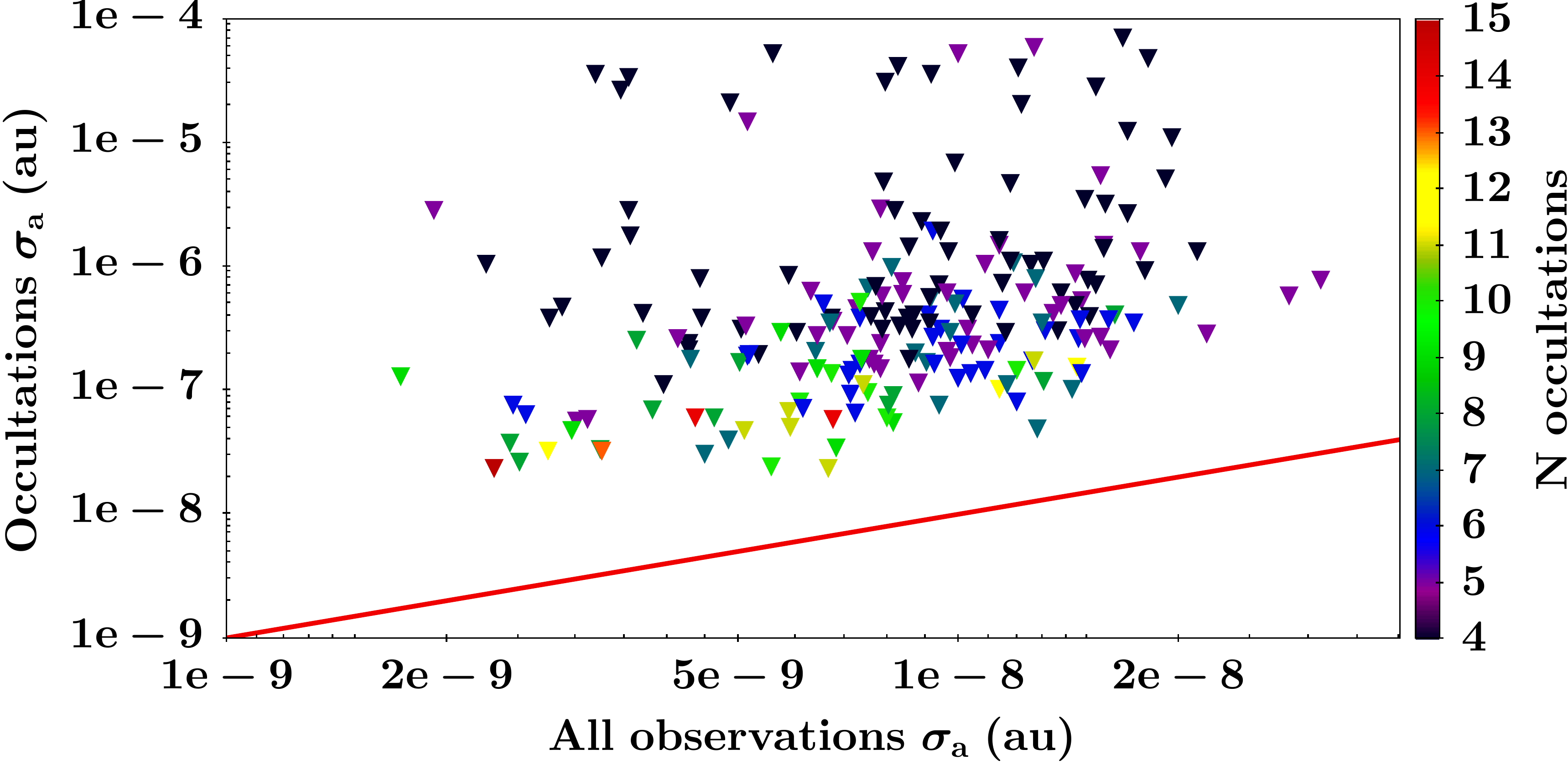}
  \caption{Upper panel: Semi-major axis uncertainties for asteroids
    with occultations. The values obtained from the use of all
    observations are on the horizontal axis. On the vertical axis, we
    show the uncertainty obtained when fitting positions derived from
    occultation astrometry only, with the procedure described in the
    text. The line represents equal uncertainties. The symbol colour
    is related to the number of occultations used. Bottom panel: same
    as upper panel, but using pre-Gaia
    astrometry.}\label{fig:orbit_sigma_GDR1}
\end{figure}

For each orbit we determine the uncertainty on the semi--major axis
$\sigma_a$ and use it as an indicator of the orbit accuracy. In
Fig.~\ref{fig:orbit_sigma_GDR1} we compare the accuracy of our
results, to that obtained from fitting all the available observation
from the Minor Planet Center: the value of $\sigma_a$ is provided by
the
AstDys \footnote{http://hamilton.dm.unipi.it/astdys/index.php?pc=0}
online repository. One should note that this last orbital solution
also contains the contributions of stellar occultations. However, as
up to now no case--by--case study of occultation astrometry has been
carried out, all such measurements were given a weight of 200 mas
\citep{Farnocchia_2015}. Such a weight, coupled to their small number
relative to approximately thousands of CCD data points, makes their role
negligible in the MPC orbital solution.

We have also computed the orbital uncertainties that could be obtained
from the original stellar positions given in the occultation
database. These are extracted, case by case, from the best data
available, mostly from Tycho/Hipparcos, UCAC2, and UCAC4. A comparison
of the upper and lower panel of Fig.~\ref{fig:orbit_sigma_GDR1} clearly
shows that GDR1 brings an overall improvement by a full order of
magnitude.

Fig.~\ref{fig:orbit_sigma_GDR1} shows that properly weighted
occultations, even if taken alone, can provide very reasonable orbital
solutions for some objects. Of course this is only true for asteroids
with a sufficient number of occultations of good quality. All the bad
orbital solutions with $\sigma_a > 10^{-6}$~au have 4-5 astrometric
positions from stellar occultations. As a rule of thumb, we can say
that above approximately ten astrometric points, uncertainties are not
worse than one order of magnitude with respect to the solution
obtained with all the observations (thousands of measurements in
general).

The performance obtained on a few asteroids, whose orbits show better
residuals when only occultations are used, is remarkable. However, for
a similar number of occultations, there is a considerable spread in
the quality of the solution from one asteroid to the other. For
instance, among the best performers, objects with 10 to 15
occultations are found, but also several with only 4 or 5 observed
events. For an interpretation of this evidence, both the occultation
quality and the distribution of the observed occultations along the
asteroid orbit have to be considered.

\begin{figure}[h]
  \centering
  \includegraphics[width=7 cm]{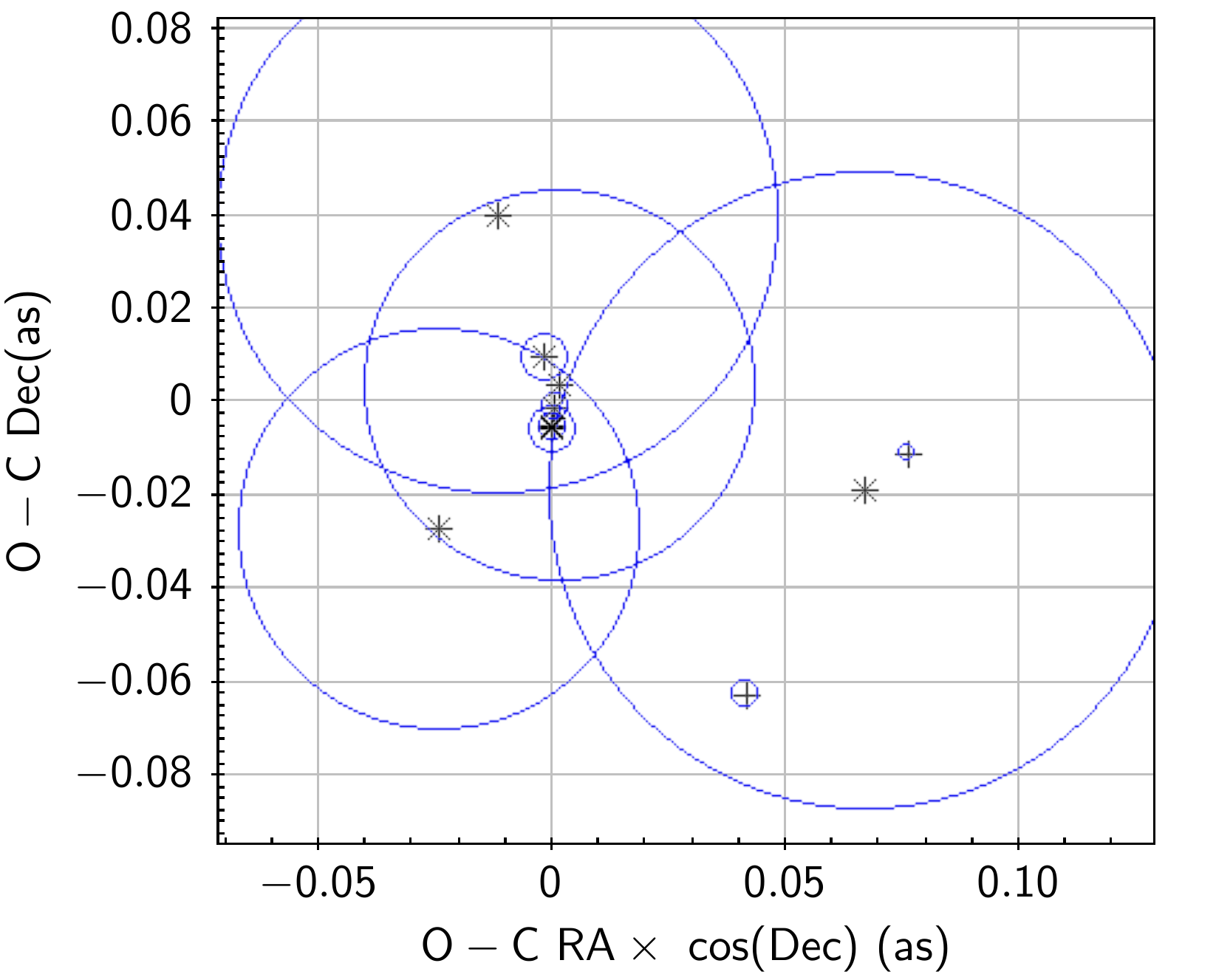}
  \includegraphics[width=7 cm]{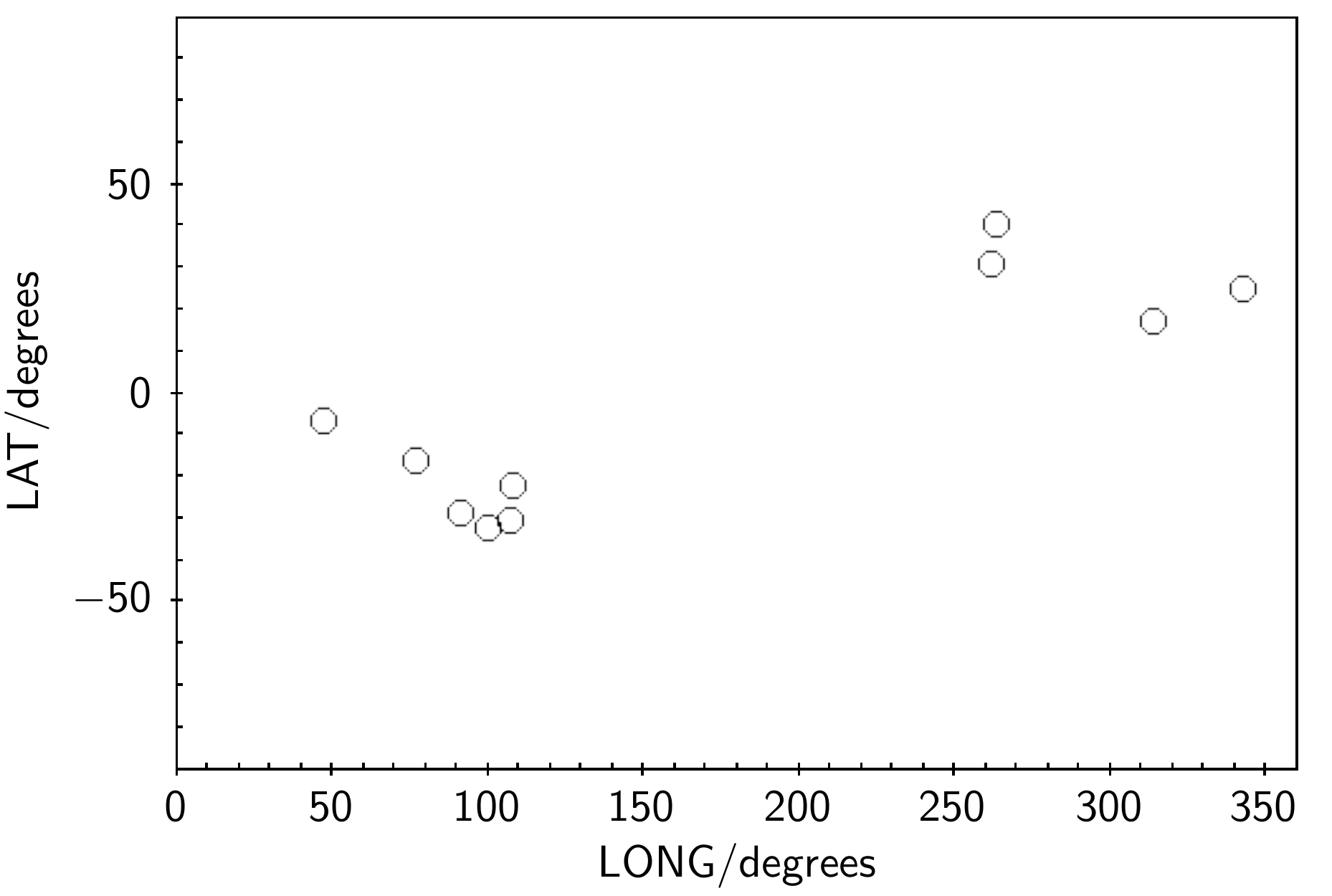}
  \caption{Upper panel: residuals, with respect to the orbit computed
    from occultations, for the astrometry of (105) Artemis. Each
    symbol represents a single occultation event. Observations that
    are used for the final orbital fit are marked with stars. Crosses
    correspond to measurements that are automatically rejected. The
    circles show the nominal uncertainty (weight) of the
    occultations. Multi--chord events correspond to smaller
    circles. In the lower panel, the distribution of the occulted
    stars in ecliptic coordinates is shown.}\label{fig:105Artemis_ecl}
\end{figure}

\begin{figure}[h]
  \centering
  \includegraphics[width=7 cm]{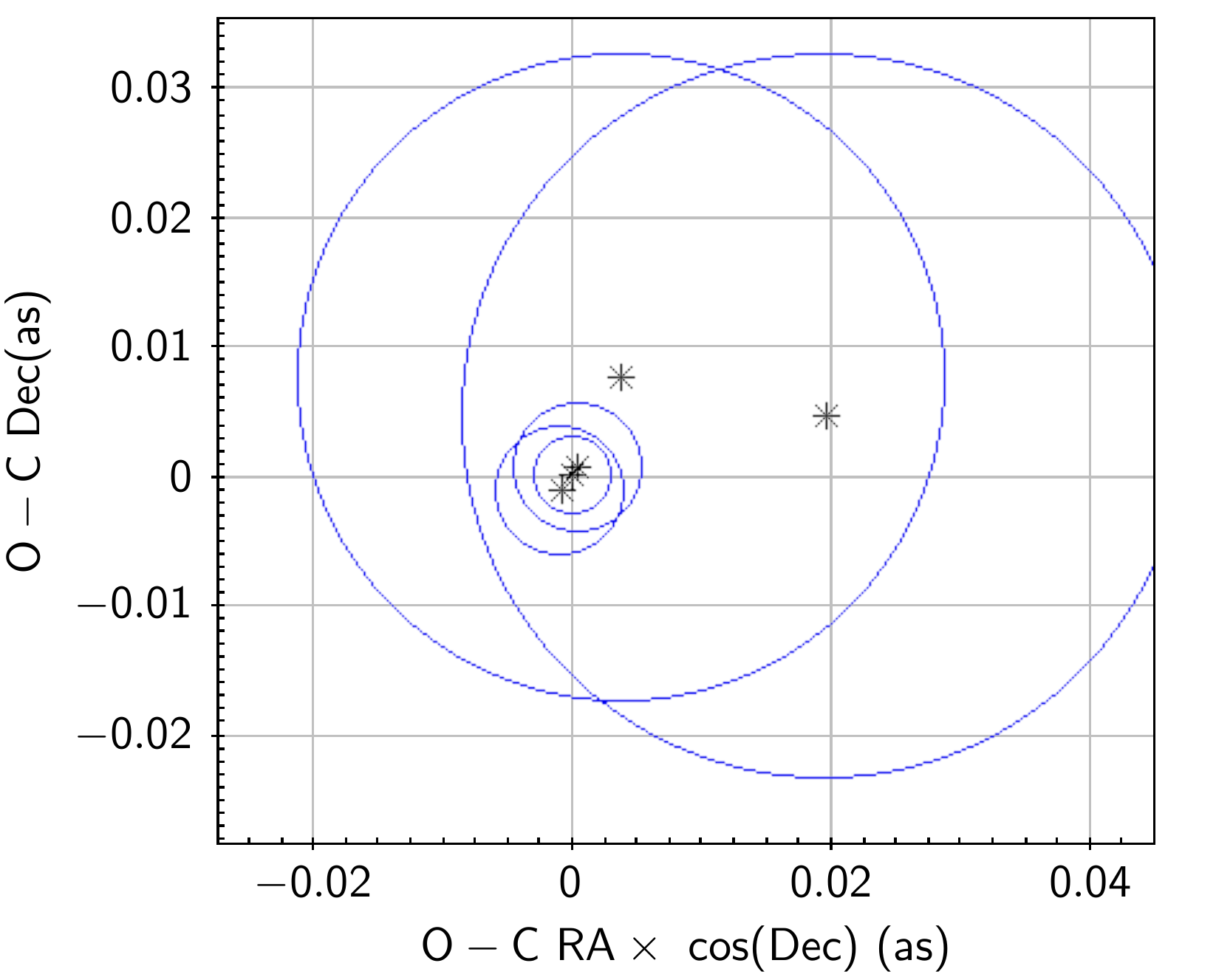}
  \includegraphics[width=7 cm]{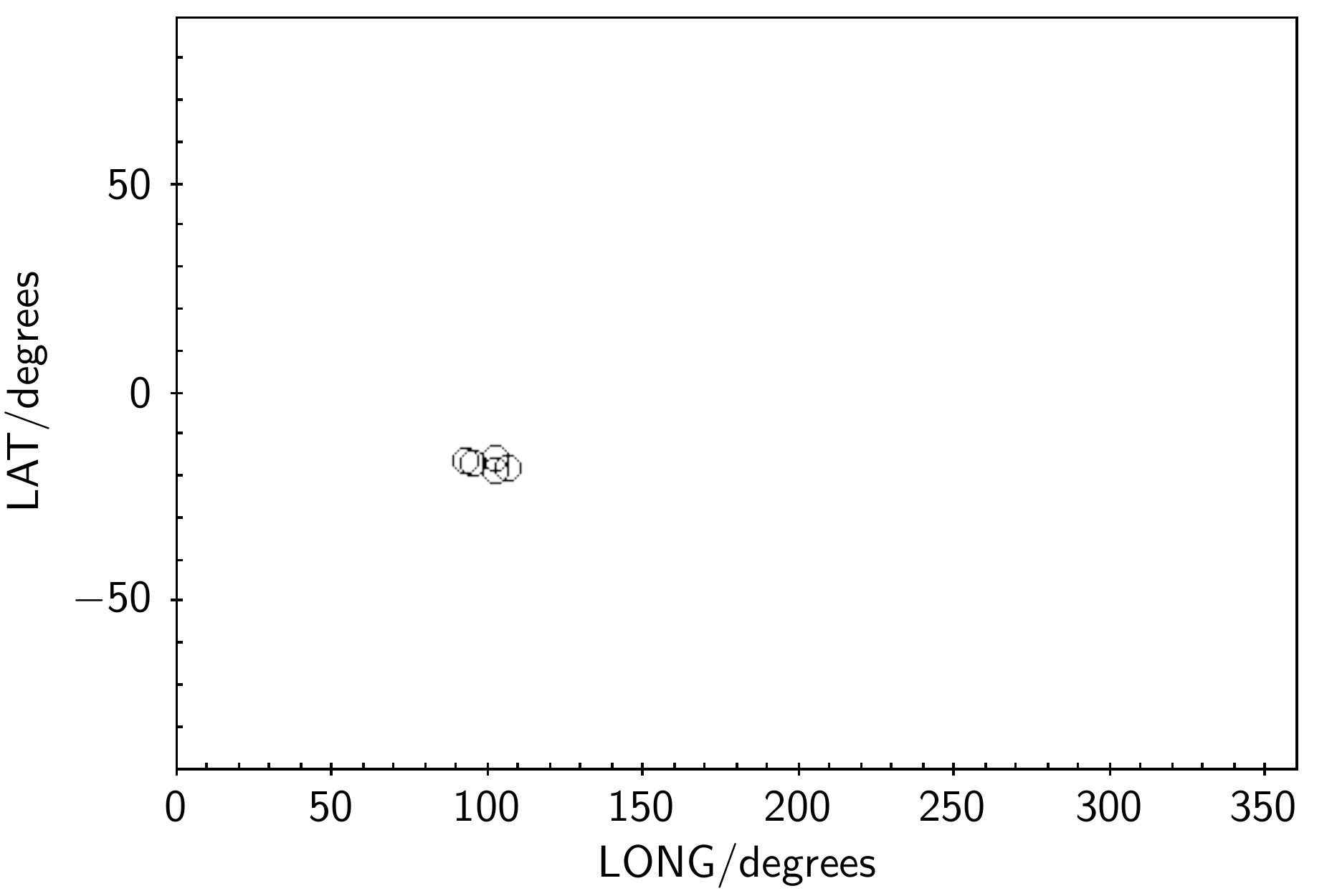}
  \caption{Same as Fig.~\ref{fig:105Artemis_ecl}, for the asteroid (234) Barbara.}\label{fig:234Barbara_ecl}
\end{figure}

We illustrate two typical situations in the following. The first one
concerns the asteroid (105) Artemis, having ten occultations covering
34 years and closely matching in performance the accuracy by
occultation astrometry and the one with ``all data'' (1733 astrometric
measurements, spanning 112 years). The orbit uncertainty is
$\sigma_{a} \sim 4 \times 10^{-9}$~au. An inspection of the residuals
of the orbit fitted to occultation astrometry
(Fig.~\ref{fig:105Artemis_ecl}, upper panel) shows that few
multi--chord events are clustered around residuals smaller than
10~mas. A couple of observations are rejected, as their residuals are
too large ($\sim$70 mas) with respect to their weight. This anomaly
could be due to specific problems with the occultations (undetected
errors in the observation or the interpretation) or with the stellar
astrometry. Eventually, four events have uncertainties comparable to the
apparent radius of the asteroid at the occultation epochs (60-70~mas)
and are fully compatible with the solution.

The corresponding distribution of the occulted stars in ecliptic
longitude is also shown and covers relatively well about two thirds of
the orbit. A good coverage in longitude is certainly required to
appropriately constrain the orbit. A counter example is provided by
the case of (234) Barbara (Fig.~\ref{fig:234Barbara_ecl}). The
residuals for the five occultations (covering less than one year) span
a smaller range (20~mas) than the case of (105) Artemis (100~mas), and
three multi-chord events exhibit residuals better than 2~mas. Yet, the
uncertainty for the occultation orbit is poor, around $\sigma_{a} \sim
5 \times 10^{-5}$~au. This is clearly explained by the fact that the
observations are strongly clustered on a very restricted arc of the
asteroid orbit, thus preventing any accurate solution.

\section{Conclusions}
\label{S:discussion}

We illustrate three situations on which the first Gaia data release,
GDR1, has an immediate impact regarding astrometry and orbits of
  SSOs: the rating of the NEA impact risk, the exploitation of data
on very short arcs, and that of stellar occultations. They
have each a factor of improvement, with respect to
previous catalogues, of about one order of magnitude, corresponding to
the ambition of GDR1.

To put our results in perspective, we expect that another dramatic
improvement will occur with GDR2, that will benefit from a much higher
number of observations and more accurate calibrations and will include
parallaxes and proper motions for all stars.

We consider that the perspective of exploiting stellar occultations
for obtaining precise astrometry is particularly interesting. In fact,
in this case the accuracy can be close to that of the star. As we
dispose of a complete record of past observations over a few decades,
and future observations will be secured in growing numbers, there
exists a concrete perspective of expanding the time frame of
Gaia-level astrometry beyond the duration of the mission.

\section*{Acknowledgements}

We thank the referee Dr. Fabrizio Bernardi for his helpful comments,
which have improved the quality of the paper.  This work has made use
of data from the European Space Agency (ESA) mission Gaia
(http://www.cosmos.esa.int/gaia), processed by the Gaia Data
Processing and Analysis Consortium (DPAC,
http://www.cosmos.esa.int/web/gaia/dpac/consortium). Funding for the
DPAC has been provided by national institutions, in particular the
institutions participating in the Gaia Multilateral Agreement.

\bibliographystyle{aa}
\bibliography{ref_aster}

\begin{thebibliography}{19}
\expandafter\ifx\csname natexlab\endcsname\relax\def\natexlab#1{#1}\fi

\bibitem[{{Altmann} {et~al.}(2014){Altmann}, {Bouquillon}, {Taris}, {Steele},
  {Smart}, {Andrei}, {Barache}, {Carlucci}, \& {Els}}]{Altmann_2014}
{Altmann}, M., {Bouquillon}, S., {Taris}, F., {et~al.} 2014, in Proceedings of
  the SPIE, Vol. 9149, Observatory Operations: Strategies, Processes, and
  Systems V, 91490P

\bibitem[{{Bouquillon} {et~al.}(2014){Bouquillon}, {Barache}, {Carlucci},
  {Taris}, {Altmann}, {Andrei}, {Smart}, {Steele}, \& {Els}}]{Bouquillon_2014}
{Bouquillon}, S., {Barache}, C., {Carlucci}, T., {et~al.} 2014, in Proceeding
  of the SPIE, Vol. 9152, Software and Cyberinfrastructure for Astronomy III,
  915203

\bibitem[{{Bouquillon} {et~al.}(2017){Bouquillon}, {Mendez}, {Altmann},
  {Barache}, {Carlucci}, {Taris}, {Andrei}, \& {Smart}}]{Bouquillon_2017}
{Bouquillon}, S., {Mendez}, R.~A., {Altmann}, M., {et~al.} 2017, accepted by
  Astronomy \& Astrophysics

\bibitem[{{Carpino} {et~al.}(2003){Carpino}, {Milani}, \&
  {Chesley}}]{Carpino_2003}
{Carpino}, M., {Milani}, A., \& {Chesley}, S.~R. 2003, Icarus, 166, 248

\bibitem[{{Chesley} {et~al.}(2010){Chesley}, {Baer}, \& {Monet}}]{Chesley_2010}
{Chesley}, S.~R., {Baer}, J., \& {Monet}, D.~G. 2010, Icarus, 210, 158

\bibitem[{Desmars {et~al.}(2013)Desmars, Bancelin, Hestroffer, \&
  Thuillot}]{desmars_statistical_2013}
Desmars, J., Bancelin, D., Hestroffer, D., \& Thuillot, W. 2013, Astronomy \&
  Astrophysics, 554, A32

\bibitem[{{Dunham} {et~al.}(2016){Dunham}, {Herald}, {Frappa}, {Hayamizu},
  {Talbot}, \& {Timerson}}]{pds_occultations_2016}
{Dunham}, D.~W., {Herald}, D., {Frappa}, E., {et~al.} 2016, NASA Planetary Data
  System, 243

\bibitem[{{Farnocchia} {et~al.}(2015{\natexlab{a}}){Farnocchia}, {Chesley},
  {Chamberlin}, \& {Tholen}}]{Farnocchia_2015}
{Farnocchia}, D., {Chesley}, S.~R., {Chamberlin}, A.~B., \& {Tholen}, D.~J.
  2015{\natexlab{a}}, Icarus, 245, 94

\bibitem[{{Farnocchia} {et~al.}(2015{\natexlab{b}}){Farnocchia}, {Chesley}, \&
  {Micheli}}]{Farnocchia2015_Systematic}
{Farnocchia}, D., {Chesley}, S.~R., \& {Micheli}, M. 2015{\natexlab{b}},
  Icarus, 258, 18

\bibitem[{{Gaia Collaboration} {et~al.}(2016){Gaia Collaboration}, Brown,
  Vallenari, Prusti, de~Bruijne, Mignard, Drimmel, Babusiaux, Bailer-Jones,
  Bastian, Biermann, Evans, Eyer, Jansen, Jordi, Katz, Klioner, Lammers,
  Lindegren, Luri, O'Mullane, Panem, Pourbaix, Randich, Sartoretti, Siddiqui,
  Soubiran, Valette, van Leeuwen, Walton, Aerts, Arenou, Cropper, Høg,
  Lattanzi, Grebel, Holland, Huc, Passot, Perryman, Bramante, Cacciari,
  Castañeda, Chaoul, Cheek, De~Angeli, Fabricius, Guerra, Hernández,
  Jean-Antoine-Piccolo, Masana, Messineo, Mowlavi, Nienartowicz,
  Ordóñez-Blanco, Panuzzo, Portell, Richards, Riello, Seabroke, Tanga,
  Thévenin, Torra, Els, Gracia-Abril, Comoretto, Garcia-Reinaldos, Lock,
  Mercier, Altmann, Andrae, Astraatmadja, Bellas-Velidis, Benson, Berthier,
  Blomme, Busso, Carry, Cellino, Clementini, Cowell, Creevey, Cuypers,
  Davidson, De~Ridder, de~Torres, Delchambre, Dell'Oro, Ducourant, Frémat,
  García-Torres, Gosset, Halbwachs, Hambly, Harrison, Hauser, Hestroffer,
  Hodgkin, Huckle, Hutton, Jasniewicz, Jordan, Kontizas, Korn, Lanzafame,
  Manteiga, Moitinho, Muinonen, Osinde, Pancino, Pauwels, Petit, Recio-Blanco,
  Robin, Sarro, Siopis, Smith, Smith, Sozzetti, Thuillot, van Reeven, Viala,
  Abbas, Abreu~Aramburu, Accart, Aguado, Allan, Allasia, Altavilla, Álvarez,
  Alves, Anderson, Andrei, Anglada~Varela, Antiche, Antoja, Antón, Arcay,
  Bach, Baker, Balaguer-Núñez, Barache, Barata, Barbier, Barblan, Barrado~y
  Navascués, Barros, Barstow, Becciani, Bellazzini, Bello~García, Belokurov,
  Bendjoya, Berihuete, Bianchi, Bienaymé, Billebaud, Blagorodnova,
  Blanco-Cuaresma, Boch, Bombrun, Borrachero, Bouquillon, Bourda, Bouy,
  Bragaglia, Breddels, Brouillet, Brüsemeister, Bucciarelli, Burgess, Burgon,
  Burlacu, Busonero, Buzzi, Caffau, Cambras, Campbell, Cancelliere,
  Cantat-Gaudin, Carlucci, Carrasco, Castellani, Charlot, Charnas, Chiavassa,
  Clotet, Cocozza, Collins, Costigan, Crifo, Cross, Crosta, Crowley, Dafonte,
  Damerdji, Dapergolas, David, David, De~Cat, de~Felice, de~Laverny, De~Luise,
  De~March, de~Martino, de~Souza, Debosscher, del Pozo, Delbo, Delgado,
  Delgado, Di~Matteo, Diakite, Distefano, Dolding, Dos~Anjos, Drazinos, Duran,
  Dzigan, Edvardsson, Enke, Evans, Eynard~Bontemps, Fabre, Fabrizio, Faigler,
  Falcão, Farràs~Casas, Federici, Fedorets, Fernández-Hernández, Fernique,
  Fienga, Figueras, Filippi, Findeisen, Fonti, Fouesneau, Fraile, Fraser,
  Fuchs, Gai, Galleti, Galluccio, Garabato, García-Sedano, Garofalo, Garralda,
  Gavras, Gerssen, Geyer, Gilmore, Girona, Giuffrida, Gomes, González-Marcos,
  González-Núñez, González-Vidal, Granvik, Guerrier, Guillout, Guiraud,
  Gúrpide, Gutiérrez-Sánchez, Guy, Haigron, Hatzidimitriou, Haywood, Heiter,
  Helmi, Hobbs, Hofmann, Holl, Holland, Hunt, Hypki, Icardi, Irwin, Jevardat~de
  Fombelle, Jofré, Jonker, Jorissen, Julbe, Karampelas, Kochoska, Kohley,
  Kolenberg, Kontizas, Koposov, Kordopatis, Koubsky, Krone-Martins,
  Kudryashova, Kull, Bachchan, Lacoste-Seris, Lanza, Lavigne,
  Le~Poncin-Lafitte, Lebreton, Lebzelter, Leccia, Leclerc, Lecoeur-Taibi,
  Lemaitre, Lenhardt, Leroux, Liao, Licata, Lindstrøm, Lister, Livanou, Lobel,
  Löffler, López, Lorenz, MacDonald, Magalhães~Fernandes, Managau, Mann,
  Mantelet, Marchal, Marchant, Marconi, Marinoni, Marrese, Marschalkó,
  Marshall, Martín-Fleitas, Martino, Mary, Matijevič, Mazeh, McMillan,
  Messina, Michalik, Millar, Miranda, Molina, Molinaro, Molinaro, Molnár,
  Moniez, Montegriffo, Mor, Mora, Morbidelli, Morel, Morgenthaler, Morris,
  Mulone, Muraveva, Musella, Narbonne, Nelemans, Nicastro, Noval, Ordénovic,
  Ordieres-Meré, Osborne, Pagani, Pagano, Pailler, Palacin, Palaversa,
  Parsons, Pecoraro, Pedrosa, Pentikäinen, Pichon, Piersimoni, Pineau, Plachy,
  Plum, Poujoulet, Prša, Pulone, Ragaini, Rago, Rambaux, Ramos-Lerate,
  Ranalli, Rauw, Read, Regibo, Reylé, Ribeiro, Rimoldini, Ripepi, Riva, Rixon,
  Roelens, Romero-Gómez, Rowell, Royer, Ruiz-Dern, Sadowski,
  Sagristà~Sellés, Sahlmann, Salgado, Salguero, Sarasso, Savietto,
  Schultheis, Sciacca, Segol, Segovia, Segransan, Shih, Smareglia, Smart,
  Solano, Solitro, Sordo, Soria~Nieto, Souchay, Spagna, Spoto, Stampa, Steele,
  Steidelmüller, Stephenson, Stoev, Suess, Süveges, Surdej, Szabados,
  Szegedi-Elek, Tapiador, Taris, Tauran, Taylor, Teixeira, Terrett, Tingley,
  Trager, Turon, Ulla, Utrilla, Valentini, van Elteren, Van~Hemelryck, van
  Leeuwen, Varadi, Vecchiato, Veljanoski, Via, Vicente, Vogt, Voss, Votruba,
  Voutsinas, Walmsley, Weiler, Weingrill, Wevers, Wyrzykowski, Yoldas, Žerjal,
  Zucker, Zurbach, Zwitter, Alecu, Allen, Allende~Prieto, Amorim,
  Anglada-Escudé, Arsenijevic, Azaz, Balm, Beck, Bernstein, Bigot, Bijaoui,
  Blasco, Bonfigli, Bono, Boudreault, Bressan, Brown, Brunet, Bunclark,
  Buonanno, Butkevich, Carret, Carrion, Chemin, Chéreau, Corcione, Darmigny,
  de~Boer, de~Teodoro, de~Zeeuw, Delle~Luche, Domingues, Dubath, Fodor,
  Frézouls, Fries, Fustes, Fyfe, Gallardo, Gallegos, Gardiol, Gebran, Gomboc,
  Gómez, Grux, Gueguen, Heyrovsky, Hoar, Iannicola, Isasi~Parache, Janotto,
  Joliet, Jonckheere, Keil, Kim, Klagyivik, Klar, Knude, Kochukhov, Kolka, Kos,
  Kutka, Lainey, LeBouquin, Liu, Loreggia, Makarov, Marseille, Martayan,
  Martinez-Rubi, Massart, Meynadier, Mignot, Munari, Nguyen, Nordlander,
  Ocvirk, O'Flaherty, Olias~Sanz, Ortiz, Osorio, Oszkiewicz, Ouzounis, Palmer,
  Park, Pasquato, Peltzer, Peralta, Péturaud, Pieniluoma, Pigozzi, Poels,
  Prat, Prod'homme, Raison, Rebordao, Risquez, Rocca-Volmerange, Rosen,
  Ruiz-Fuertes, Russo, Sembay, Serraller~Vizcaino, Short, Siebert, Silva,
  Sinachopoulos, Slezak, Soffel, Sosnowska, Straižys, ter Linden, Terrell,
  Theil, Tiede, Troisi, Tsalmantza, Tur, Vaccari, Vachier, Valles, Van~Hamme,
  Veltz, Virtanen, Wallut, Wichmann, Wilkinson, Ziaeepour, \&
  Zschocke}]{gaia_collaboration_gaia_2016}
{Gaia Collaboration}, Brown, A. G.~A., Vallenari, A., {et~al.} 2016, Astronomy
  and Astrophysics, 595, A2

\bibitem[{{Liebe}(1995)}]{Liebe_1995}
{Liebe}, C.~C. 1995, IEEE Aerospace and Electronic Systems Magazine, 10, 10

\bibitem[{{Lindegren, L.} {et~al.}(2016){Lindegren, L.}, {Lammers, U.},
  {Bastian, U.}, {Hernández, J.}, {Klioner, S.}, {Hobbs, D.}, {Bombrun, A.},
  {Michalik, D.}, {Ramos-Lerate, M.}, {Butkevich, A.}, {Comoretto, G.},
  {Joliet, E.}, {Holl, B.}, {Hutton, A.}, {Parsons, P.}, {Steidelmüller, H.},
  {Abbas, U.}, {Altmann, M.}, {Andrei, A.}, {Anton, S.}, {Bach, N.}, {Barache,
  C.}, {Becciani, U.}, {Berthier, J.}, {Bianchi, L.}, {Biermann, M.},
  {Bouquillon, S.}, {Bourda, G.}, {Brüsemeister, T.}, {Bucciarelli, B.},
  {Busonero, D.}, {Carlucci, T.}, {Castañeda, J.}, {Charlot, P.}, {Clotet,
  M.}, {Crosta, M.}, {Davidson, M.}, {de Felice, F.}, {Drimmel, R.},
  {Fabricius, C.}, {Fienga, A.}, {Figueras, F.}, {Fraile, E.}, {Gai, M.},
  {Garralda, N.}, {Geyer, R.}, {González-Vidal, J. J.}, {Guerra, R.}, {Hambly,
  N. C.}, {Hauser, M.}, {Jordan, S.}, {Lattanzi, M. G.}, {Lenhardt, H.}, {Liao,
  S.}, {Löffler, W.}, {McMillan, P. J.}, {Mignard, F.}, {Mora, A.},
  {Morbidelli, R.}, {Portell, J.}, {Riva, A.}, {Sarasso, M.}, {Serraller, I.},
  {Siddiqui, H.}, {Smart, R.}, {Spagna, A.}, {Stampa, U.}, {Steele, I.},
  {Taris, F.}, {Torra, J.}, {van Reeven, W.}, {Vecchiato, A.}, {Zschocke, S.},
  {de Bruijne, J.}, {Gracia, G.}, {Raison, F.}, {Lister, T.}, {Marchant, J.},
  {Messineo, R.}, {Soffel, M.}, {Osorio, J.}, {de Torres, A.}, \& {O’Mullane,
  W.}}]{gaia_astrometry_2016}
{Lindegren, L.}, {Lammers, U.}, {Bastian, U.}, {et~al.} 2016, Astronomy and
  Astrophysics, 595, A4

\bibitem[{{Milani} {et~al.}(2005{\natexlab{a}}){Milani}, {Chesley},
  {Sansaturio}, {Tommei}, \& {Valsecchi}}]{Milani_2005}
{Milani}, A., {Chesley}, S.~R., {Sansaturio}, M.~E., {Tommei}, G., \&
  {Valsecchi}, G.~B. 2005{\natexlab{a}}, Icarus, 173, 362

\bibitem[{{Milani} {et~al.}(2004){Milani}, {Gronchi}, {Vitturi}, \& {Kne{\v
  z}evi{\'c}}}]{Milani_2005_admissible}
{Milani}, A., {Gronchi}, G.~F., {Vitturi}, M.~D., \& {Kne{\v z}evi{\'c}}, Z.
  2004, Celestial Mechanics and Dynamical Astronomy, 90, 57

\bibitem[{{Milani} {et~al.}(2005{\natexlab{b}}){Milani}, {Sansaturio},
  {Tommei}, {Arratia}, \& {Chesley}}]{Milani_2005_metrics}
{Milani}, A., {Sansaturio}, M.~E., {Tommei}, G., {Arratia}, O., \& {Chesley},
  S.~R. 2005{\natexlab{b}}, Astronomy \& Astrophysics, 431, 729

\bibitem[{{Roeser} {et~al.}(2010){Roeser}, {Demleitner}, \&
  {Schilbach}}]{Roeser_2010}
{Roeser}, S., {Demleitner}, M., \& {Schilbach}, E. 2010, Astronomical Journal,
  139, 2440

\bibitem[{{Spoto} {et~al.}(2017){Spoto}, {Milani}, {Tommei}, {Del Vigna},
  {Tanga}, \& {Mignard}}]{Spoto2017}
{Spoto}, F., {Milani}, A., {Tommei}, G., {et~al.} 2017, To be submitted

\bibitem[{Tanga \& Delbo(2007)}]{tanga_asteroid_2007}
Tanga, P. \& Delbo, M. 2007, Astronomy and Astrophysics, 474, 1015

\bibitem[{Tanga {et~al.}(2008)Tanga, Hestroffer, Delbò, Frouard, Mouret, \&
  Thuillot}]{tanga_gaia_2008}
Tanga, P., Hestroffer, D., Delbò, M., {et~al.} 2008, Planetary and Space
  Science, 56, 1812

\end{thebibliography}

\end{document}